\documentstyle[12pt,A4]{article}

\begin{document}
\label{vys1-zac}
\noindent
\hspace*{8cm}{\sc acta univ. palacki. olomuc.,}\\
\hspace*{8cm}{\sc fac.\,rer.\,nat.\,{\small (1997)},\,physica\,{\small
36}},\\
\hspace*{8cm}{\small \pageref{vys1-zac} -- \pageref{vys1-kon}}\\
\hspace*{8cm}\rule[3mm]{6.2cm}{0.2mm}\\[1cm]
\centerline{\large \bf OPTICAL ROTATORY DISPERSION OF $\alpha
$\,-\,QUARTZ}\\[2mm]
\centerline{\bf Vratislav Vy\v s\'\i n, Ivo Vy\v s\'\i n}\\[3mm]
\noindent {\small Department of Theoretical Physics, Natural Science
Faculty of Palack˜ University, Svobody 26, {\bf 771~46 Olomouc}, Czech
Republic}\\[4mm]
\centerline{\it Received 26th September 1996}

\vspace*{0.5cm}

{\sc ABSTRACT:} It is shown that some formulae describing optical rotatory dispersion of
$\alpha$-quartz with the aid of two Drude's terms reduce to the
combined formula containing one Drude's and one Chandrasekhar's term.
Comparison of various formulae describing the experimental data of
$\alpha$-quartz leads to the conclusion that the optical activity of
this crystal is due to its crystal structure only, that means the
optical activity is not of molecular origin. Further the rotatory
strengths are discussed with the regard to coupled oscillator model and
to the structure of $\alpha$-quartz.

\section{Introduction} \label{int}
\hspace \parindent
Burkov et alii \cite{1,2} have undertaken an exhaustive study
of formulae describing optical rotatory dispersion (ORD) of
$\alpha$-quartz. This study is based on precise measurements of various
authors \cite{3,4,5} in the region from $0.15 \mu m$ to $3.2 \mu m$. Beside
the well known formula of Drude
\begin{equation}
\rho (\lambda )=\sum_{i}\frac{K_{i}^{(1)}}{\lambda ^{2}-\lambda
_{i}^{2}}
\label{1}
\end{equation}
Burkov et alii have used the following formulae:
\begin{equation}
\rho (\lambda )=\sum_{i}\frac{K_{i}^{(1)}\lambda ^{2}}{\left(\lambda
^{2}-\lambda_{i}^{2}\right)^{2}}
\label{2}
\end{equation}
derived by Chandrasekhar \cite{6,7}, the combined formula
\begin{equation}
\rho (\lambda )=\sum_{i}\left[\frac{K_{i}^{(1)}}{\lambda ^{2}-\lambda
_{i}^{2}}+\frac{K_{i}^{(2)}\lambda ^{2}}{\left(\lambda ^{2}-\lambda
_{i}^{2}\right)^{2}}\right]
\label{3}
\end{equation}
derived by one of us \cite{8} and also by Nelson \cite{8a} and formula
\begin{equation}
\rho (\lambda )=\sum_{i}\frac{K_{i}^{(3)}\left(\lambda ^{2}+\lambda
_{i}^{2}\right)}{\left(\lambda ^{2}-\lambda _{i}^{2}\right)^{2}}
\label{4}
\end{equation}
derived by Agranovich \cite{9,10} and by Tsvirko \cite{11}. In these formulae
$\rho$ is the rotatory power, $\lambda _{i}$ is the characteristic
wavelength and $K_{i}^{(j)}$ are the constants including molecular and
crystalline properties of $\alpha $-quartz. Formulae (\ref{2}) and
(\ref{3}) are
based on the theory of coupled oscillators and the formula (\ref{4}) is based on
the excitons theory. It should be noted that measurements and theories
for light propagating along the optic axis have been taken into account
only. The applicability of the above mentioned formulae have been judged by
Burkov et alii with the aid of root-mean-square deviation defined by
\begin{equation}
g=\sum\limits _{n}(\rho _{exp}-\rho _{teor})^{2}
\label{5}
\end{equation}
where $\rho _{exp}$ are experimental data and $\rho _{teor}$ fitted data
following from formulae (\ref{1}) - (\ref{4}). Further $n$ is a number
of measured
data and the analysis of Burkov et alii comprises forty-five points from
$0.15\mu m$ to $3.2\mu m$. These authors have found that formulae
(\ref{1})
and (\ref{4}) with one term are inconvenient because $g$ is very great.
They have
found good agreement in the case of formula (\ref{1}) with two terms of
opposite signs, in the case of formula (\ref{2}) with one or two terms and in
the case of formula (\ref{3}) with different $K_{1}^{(1)}$ and $K_{2}^{(2)}$
but with one $\lambda _{i}$. In some cases they have introduced the
constant term corresponding to ORD in infra-red region. The results of
the most convenient formulae are presented in Tab. \ref{tab1}.

\begin{table}[ht]
\caption{Table of formulae.}
\begin{center}
\begin{tabular}{|r|l|c|}
\hline
Nr. & Formula & $g$\\[1mm]
\hline
I & $\rho (\lambda )=\frac{7.17\lambda ^{2}}{\left(\lambda
^{2}-0.0928^{2}\right)^{2}}$ & 3.91\\[2mm]
II & $\rho (\lambda )=\frac{-0.05}{\lambda
^{2}-0.09266^{2}}+\frac{7.266\lambda ^{2}}{\left(\lambda
^{2}-0.09266^{2}\right)^{2}}$ & 3.90\\[2mm]
III & $\rho (\lambda )=\frac{184.22}{\lambda
^{2}-0.09355^{2}}+\frac{-177.04}{\lambda ^{2}-0.09170^{2}}$ &
3.90\\[2mm]
IV & $\rho (\lambda )=\frac{-1.05\lambda ^{2}}{\left(\lambda
^{2}-0.09998^{2}\right)^{2}}+\frac{8.22\lambda ^{2}}{\left(\lambda
^{2}-0.09387^{2}\right)^{2}}$ & 3.82\\[2mm]
V & $\rho (\lambda )=\frac{219.47}{\lambda
^{2}-0.09332^{2}}+\frac{-212.30}{\lambda ^{2}-0.09176^{2}}+0.0361$ &
3.89\\[2mm]
VI & $\rho (\lambda )=\frac{-1.335\lambda ^{2}}{\left(\lambda
^{2}-0.1012^{2}\right)^{2}}+\frac{8.49\lambda ^{2}}{\left(\lambda
^{2}-0.0944^{2}\right)^{2}}+0.0964$ & 3.74\\[2mm]
\hline
\end{tabular}
\label{tab1}
\end{center}
\end{table}

In addition it should be noted that Katzin and B\H urer \cite{12,13} have
worked out a formula for ORD of $\alpha $-quartz in the range from
$0.23\mu m$ to $3.5\mu m$. Their formula has the form
\begin{equation}
\rho (\lambda )=\frac{127.02476}{\lambda
^{2}-0.0979^{2}}-\frac{119.77145}{\lambda ^{2}-0.0958^{2}}-0.1879.
\label{6}
\end{equation}

The experimental data are fitted to a root-mean-square deviation of
about $0.038deg$. It is the most accurate formula worked out for this
spectral range. However, this formula deviates notably from important
measurements of Servant \cite{5} in the far ultraviolet. In the range from
$0.1850\mu m$ to $0.15235\mu m$ the deviations are of systematic nature
and become very great. E. g. for $\lambda =0.15235\mu m$ rotatory power
is $779.9deg/mm$ whereas formula (6) gives $791.7deg/mm$.

The good agreement of formulae containing the wave dependence $\lambda
^{2}/(\lambda ^{2}-\lambda _{i}^{2})^{2}$ is not surprising because this
dependence holds for some other crystals which are, like $\alpha
$-quartz, optically active in crystalline state only. On the other
hand good approximation with the aid of formulae containing Drude's
terms only such as the formulae III and V from Tab. \ref{tab1} is surprising because we are
suspicious, in agreement with Hennessey and Vedam \cite{14}, of the formulae
containing Drude's term only. These terms are typical for optically
active molecules. The molecular origin of optical activity of $\alpha
$-quartz is highly improbable because of the high symmetry of $SiO_{2}$
groups. Of course, they are some crystals the ORD of which can be fitted
by Drude's formula although they consist from symmetrical
molecules. In this case the optical activity is of a pseudomolecular
origin. The molecules of these crystals are symmetrical when isolated
but they lose their symmetry in the crystalline state. They play then the
same role as chromophors in optically active molecules. But it seems
that this situation occurs in the case when crystals are composed from
large molecules. We shall see that the loss of symmetry of $SiO_{2}$
groups in $\alpha $-quartz is very little. We shall show that the
formulae containing only Drude's terms reduce to our formula (\ref{3}), where
Chandrasekhar's terms plays a predominant role and Drude's terms
represents a little contribution. It means that formulae III and 
V~differs very little from formula II.

\section{Coupled Oscillators Model} \label{coupled}
\hspace \parindent
ORD of $\alpha $-quartz via coupled oscillators model has been firstly
studied by Chandrasekhar \cite{6,7} and then by us \cite{8}. This model has been
also generalized to the case of absorption region, where circular
dichroism occurs \cite{15,16}. In this section we shall improve this model
in order to obtain equations containing explicitly the rotational strengths
for normal modes of vibrations which are the specific features of this
model. We shall restrict our considerations to the crystals structure
belonging to the space group of symmetry $D_{3}^{4}$ or its enantiomer~
$D_{3}^{6}$ because $\alpha $-quartz is its typical representative.
The crystal structure of $\alpha $-quartz \cite{17,18} shows that an axial
channel which results in a lower specific gravity than would be found
from closer packing of the atoms. Fusion, rather than giving a higher
density from closer packing, results in an even lower specific gravity.
This implies a molecular $SiO_{2}$ structural unit rather than an ionic
crystalline constitution \cite{17,18}. Unit cell contains three molecules
which are arranged spirally about a set of parallel $c$-axes. This
arrangement together with the interaction between molecules causes a
gyratory effect. In order to describe the optical properties of medium under
the consideration it is natural to start from the properties of an isolated
molecule and interactions between molecules are neglected because they
play the role of little contributions. But it is not possible in the study
of the optical activity of crystals consisting from inactive molecules.
Therefore the above mentioned chiral structure together with the
interaction between
molecules must be simultaneously taken into account. In the study of
interaction of an isolated molecule with an electromagnetic radiation each
molecule behaves like a linear harmonic oscillator by which it may be
represented. In the study of the optical activity the interaction between
molecules can be considered with the aid of a compound oscillator
consisting from two linear harmonic oscillators coupled together.

According to crystal structure of $\alpha $-quartz let the first
single oscillator, representing an isolated $SiO_{2}$ group, be situated
at the origin of co-ordinate system with its vibration direction along
$\alpha ,\,\beta ,\,\gamma ,$ where $\alpha ,\,\beta ,\,\gamma $ are the
direction cosines. Each oscillator is a charged particle which is assumed
to be bound elastically to its own equilibrium position and capable of
vibrating along a line. In the case of $\alpha $-quartz we assume the
vibrations along the line joining the centers of gravity of oxygen
atoms. The second oscillator is situated at $x,\,y,\,z$ with its
vibration direction turned through an angel $\theta $ about $c$-axis with
respect to the first one. For $\alpha $-quartz $\theta =120deg$. Each
oscillator, when uncoupled, has the natural frequency $\omega _{0}$. As
a result of the coupling, $\omega _{0}$ would be split into two normal
frequencies $\omega _{1}$ and $\omega _{2}$. These normal frequencies of
the compound oscillator are
\begin{equation}
\omega _{1}=\omega _{0}^{2}+2\pi ^{2}\epsilon, \qquad \qquad \omega
_{2}^{2}=\omega _{0}^{2}-2\pi ^{2}\epsilon.
\label{7}
\end{equation}

Here $\epsilon $ is a coupling constant. The exact nature of the
coupling is not essential in the following considerations but we
consider it as small. We can neglect all terms containing $\epsilon
^{2}$. In our case it means that we assume the interactions of valence
shall electrons of each isolated $SiO_{2}$ group.

The above mentioned model enables us to evaluate $n_{l}$ and $n_{r}$ as
it has been performed by Chandrasekhar \cite{6,7}. He has assumed that both
normal modes have the same oscillator strengths. We have removed this
simplification \cite{8} obtaining for rotatory power $\rho $ the following
expression
\begin{eqnarray}
\rho (\omega )&=&\frac{\omega }{2c}\left(n_{l}-n_{r}\right)=\nonumber \\
&=&\frac{\pi Ne^{2}\left(\alpha ^{2}+\beta ^{2}\right)d\sin \theta
}{mc^{2}} \cdot \left(\frac{f_{q_{2}}\omega ^{2}}{\omega _{2}^{2}-\omega
^{2}}-\frac{f_{q_{1}}\omega ^{2}}{\omega _{1}^{2}-\omega ^{2}}\right)
\label{8}
\end{eqnarray}
or in wave lengths
\begin{equation}
\rho (\lambda )=\frac{\pi Ne^{2}\left(\alpha ^{2}+\beta ^{2}\right)d\sin
\theta }{mc^{2}}\cdot \left(\frac{f_{q_{2}}\lambda _{2}^{2}}{\lambda
^{2}-\lambda _{2}^{2}}-\frac{f_{q_{1}}\lambda _{1}^{2}}{\lambda
^{2}-\lambda _{1}^{2}}\right).
\label{9}
\end{equation}

From eqs. (\ref{7}) and (\ref{8}) one gets
\begin{equation}
\rho (\lambda )=\frac{\pi Ne^{2}\left(\alpha ^{2}+\beta ^{2}\right)d\sin
\theta }{mc^{2}}\cdot \left[\frac{\epsilon
\left(f_{q_{1}}+f_{q_{2}}\right)\lambda _{0}^{4}\lambda
^{2}}{2c^{2}\left(\lambda ^{2}-\lambda
_{0}^{2}\right)^{2}}+\frac{\left(f_{q_{2}}-f_{q_{1}}\right)\lambda
_{0}^{2}}{\lambda ^{2}-\lambda _{0}^{2}}\right].
\label{10}
\end{equation}

Here $N$ is a number of single oscillators per unit volume, $d$ is a
linear distance between two single oscillators and $f_{q_{1}}$,
$f_{q_{2}}$ are oscillator strengths of both normal modes of vibrations.

We see that this model does not contain rotational strengths explicitly.
It is because this model is a classical one. The rotational strengths
follows from the quantum mechanical equations of motion of coupled
oscillators. It will be shown in Appendix. Each normal mode has a proper
rotational strengths. The strengths are
\begin{displaymath}
R_{q_{1}}=-\frac{\hbar e^{2}d\left(\alpha ^{2}+\beta
^{2}\right)f_{q_{1}}\sin \theta}{8mc}
\end{displaymath}
\begin{equation}
R_{q_{2}}=\frac{\hbar e^{2}d\left(\alpha ^{2}+\beta ^{2}\right)
f_{q_{2}}\sin \theta}{8mc}.
\label{11}
\end{equation}

Substituting from eq. (\ref{11}) into eq. (\ref{9}) we have
\begin{equation}
\rho (\lambda )=\frac{8\pi N}{\hbar c}\left(\frac{R_{q_{2}}\lambda
_{2}^{2}}{\lambda ^{2}-\lambda _{2}^{2}}+\frac{R_{q_{1}}\lambda
_{1}^{2}}{\lambda ^{2}-\lambda _{1}^{2}}\right)
\label{12}
\end{equation}
or with the aid of eq. (\ref{7})
\begin{eqnarray}
\rho (\lambda )&=&\frac{4\pi N\epsilon
\left(R_{q_{2}}-R_{q_{1}}\right)\lambda _{0}^{4}}{\hbar c^{3}}\cdot
\frac{\lambda ^{2}}{\left(\lambda ^{2}-\lambda _{0}^{2}\right)^{2}}+
\nonumber \\
&&+\frac{8\pi N\left(R_{q_{1}}+R_{q_{2}}\right)\lambda _{0}^{2}}{\hbar
c}\cdot \frac{1}{\lambda ^{2}-\lambda _{0}^{2}}.
\label{13}
\end{eqnarray}

It is clear that eqs. (\ref{12}) and (\ref{3}) are identical but under the
condition that the difference between $\lambda _{1}$ and $\lambda
_{2}$ is very small. This little difference follows from the assumption
that $\epsilon $ is small. Now we see that eq. (\ref{13}) contains two terms
the first of which is of Chandrasekhar's type while the second one is of
Drude's type. In the case of $f_{q_{1}}=f_{q_{2}}$ we have
$R_{q_{1}}=-R_{q_{2}}$ and Drude's term in eq. (\ref{13}) vanishes. Comparing
eqs. (\ref{12}) and (\ref{13}) with the formulae in Tab. \ref{tab1} we see that
formulae III
and V, being of Drude's type, contain $\lambda _{1}$ and $\lambda _{2}$
which are very near and that the constants are also very near in
numerical values but of opposite signs. We can also expect that these
formulae reduce to our combined formula (\ref{3}). Let us suppose that both
constants in formulae of the type of III and V are the same in numerical
values but with opposite signs, then these formulae reduce to
Chandrasekhar's one. We can also expect that formulae III and V reduce
to the formula which should be very near to formula II of Tab.
\ref{tab1}. We
shall see in the next section that this is true.

\section{Discussion of Formulae}
\hspace \parindent
Let us compare formula III with eq. (\ref{12}). We rewrite eq. (\ref{12})
as follows
\begin{equation}
\rho (\lambda )=\frac{A_{q_{2}}\lambda _{2}^{2}}{\lambda ^{2}-\lambda
_{2}^{2}}+\frac{A_{q_{1}}\lambda _{1}^{2}}{\lambda ^{2}-\lambda
_{1}^{2}}.
\label{14}
\end{equation}

We must take care of units. In formulae I - IV $\rho $ is expressed in
$deg/mm$ and in theoretical considerations we express it in $rad/cm$. In
eq. (\ref{14}) $\rho $ and $A_{q_{2}}$ and $A_{q_{1}}$ are in the same units.
Comparing eqs. (\ref{12}) and (\ref{14}) we see that
\begin{equation}
A_{q_{1,2}}=\frac{8\pi NR_{q_{1,2}}}{\hbar c}.
\label{15}
\end{equation}

Substituting from eq. (\ref{15}) for $R_{q_{1}}-R_{q_{2}}$ and for
$R_{q_{1}}+R_{q_{2}}$ into eq. (\ref{13}) one gets
\begin{equation}
\rho (\lambda )=\frac{\left(\epsilon \lambda
_{0}^{4}/2c^{2}\right)\left(A_{q_{2}}-A_{q_{1}}\right)\lambda
^{2}}{\left(\lambda ^{2}-\lambda
_{0}^{2}\right)^{2}}+\frac{\left(A_{q_{1}}+A_{q_{2}}\right)\lambda
_{0}^{2}}{\lambda ^{2}-\lambda _{0}^{2}},
\label{16}
\end{equation}
where $\lambda _{0}=(\lambda _{1}+\lambda _{2})/2$. Using eq. (\ref{7}) we get
\begin{equation}
\frac{\epsilon \lambda _{0}^{4}}{2c^{2}}=\frac{1}{2}\left(\lambda
_{2}^{2}-\lambda _{1}^{2}\right)=\lambda _{0}\Delta \lambda.
\label{17}
\end{equation}

Here $\Delta \lambda =\lambda _{2}-\lambda _{1}$. It follows from
formula III that $A_{q_{2}}=2.1049\cdot 10^{4}$ and
$A_{q_{1}}=-2.1053\cdot 10^{4}$ in units of $\rho $. Further $\lambda
_{0}=0.09265 \mu m$ and $\Delta \lambda =1.85 nm$. Substituting these
values into eq. (\ref{16}) we have
\begin{equation}
\rho (\lambda )=\frac{7.214\lambda ^{2}}{\left(\lambda
^{2}-0.09265^{2}\right)^{2}}-\frac{0.034}{\lambda ^{2}-0.09265^{2}}.
\label{18}
\end{equation}

In a similar way we obtain instead of formula V
\begin{equation}
\rho (\lambda )=\frac{7.281\lambda ^{2}}{\left(\lambda
^{2}-0.09254^{2}\right)^{2}}-\frac{0.013}{\lambda
^{2}-0.09254^{2}}+0.036.
\label{19}
\end{equation}

In this case $\lambda _{0}=0.09254\mu m$ and $\Delta \lambda =1.54nm$.
We are able to evaluate $\epsilon $ from eq. (\ref{17}). It is $4.25\cdot
10^{29}sec^{-2}$ for formula III and $3.54\cdot10^{29}sec^{-2}$ for
formula V. Because $\omega _{0}^{2}$ is of order $10^{32}sec^{-2}$ we
see that $\epsilon $ is really small as we have assumed. Chandrasekhar
\cite{6} has also evaluated  $\epsilon $ for $\alpha $-quartz but he had
to use beside his formula I the formula for ordinary refractive dispersion.
His value is $6.35\cdot 10^{29}sec^{-2}$.

We see that there is a little difference among formulae I, II, III and V.
We conclude that formulae III and V are nothing but formula II.

Now let us discuss the rotational strengths which may be evaluated from
eqs. (\ref{12}) and (\ref{14}). In this case we must multiply both sides
of eq. (\ref{14})
by $\pi /18$. Then we obtain $\rho $ directly in $rad/cm$, that is
\begin{equation}
\rho (\lambda )=\frac{3.664\cdot 10^{3}\lambda _{2}^{2}}{\lambda
^{2}-\lambda _{2}^{2}}-\frac{3.674\cdot 10^{3}\lambda _{1}^{2}}{\lambda
^{2}-\lambda _{1}^{2}}
\label{20}
\end{equation}
or
\begin{equation}
R_{q_{1}}=-\frac{3.674\cdot 10^{3}\hbar c}{8\pi N},\qquad \qquad
R_{q_{2}}=\frac{3.664\cdot 10^{3}\hbar c}{8\pi N}.
\label{21}
\end{equation}

We have for $\alpha $-quartz $N=2.68\cdot 10^{22}$ and therefore
$R_{q_{1}}=-1.732\cdot 10^{-37}$ and $R_{q_{2}}=1.727\cdot 10^{-37}$ in
cgs units.

We are now able to evaluate oscillator strengths $f_{q_{1}}$ and
$f_{q_{2}}$. We have for $\alpha $-quartz \cite{6} $\alpha ^{2}=0.64$, $\beta
^{2}=0.0232$, $d=1.8\cdot 10^{-28}cm$, $\sin \theta =\sqrt{3}/2$. We get
from eq. (\ref{11})
\begin{displaymath}
R_{q_{1}}=-1.155\cdot 10^{-38}f_{q_{1}}=-1.732\cdot 10^{-37}
\end{displaymath}
\begin{equation}
R_{q_{2}}=1.155\cdot 10^{-38}f_{q_{2}}=1.727\cdot 10^{-37}.
\label{22}
\end{equation}

From thus $f_{q_{1}}=14.99$ and $f_{q_{2}}=14.95$ and the average value
$f_{0}=(f_{q_{1}}+f_{q_{2}})/2=14.97$. This actual value of oscillator
strengths corresponds to the number of valence~-~shell electrons of the
oxygen and silica bring into $SiO_{2}$ molecule.

Using the same procedure in the case of formula V we get for
$R_{q_{1}}=-2.0675\cdot 10^{-37}$ and for $R_{q_{2}}=2.0677\cdot
10^{-37}$. The splitting of oscillator strengths is very little because
$f_{q_{1}}=18.852$ and $f_{q_{2}}=18.851$. For these reasons Drude's
term plays a negligible role in formula V. The averaged value of
$f_{0}=18.8515$ seems to be in high regard to the number of
valence~-~shell electrons in $SiO_{2}$. Chandrasekhar \cite{6} combining
formulae for ORD and refractive dispersion has obtained
\begin{equation}
\frac{\alpha ^{2}+\beta ^{2}}{2}f_{0}=6.5.
\label{23}
\end{equation}

From thus $f_{0}=19.68$ which is also higher then the valence-shell
electrons in $SiO_{2}$.

\section{Conclusions}
\hspace \parindent
Performing the analysis we may conclude that all formulae from Tab.
\ref{tab1} reduce to
formula in which the term $\lambda ^{2}/(\lambda ^{2}-\lambda
_{0}^{2})^{2}$ plays a predominant role. This wave dependence arises
from the model of coupled oscillators. Although this model is not
sufficiently general to be rigorously valid for a crystal, it brings out
clearly a type of interactions which may be important in determining
ORD of crystals which are optically active in crystalline state only. It
has been shown that formulae containing two Drude's terms with opposite
signs and involving $\lambda _{1}$ and $\lambda _{2}$ which are close to
each other reduce in fact to combined formula in which Chandrasekhar's
term plays a predominant role.

Nelson \cite{8a} has analysed the mechanism of crystalline optical
activity and concluded that the quadratic term in formula II is due to
lattice dynamics, while the electric-dipole--magnetic-dipole or
electric-dipole--electric-quadrupole interference leads to the known
Drude's term.

The evaluated rotational strengths of both modes of vibrations are of an
order $10^{-37}$. This value is acceptable for crystals in visible and
in uv region. On the other hand it seems that only the formula III and
for these reasons also the formula II gives the actual value of $f_{0}$
corresponding to the number of valence~-~shell electrons in $SiO_{2}$.

The location of $\lambda _{0}$ is also of great importance and
facilitates to understate the mechanism of ORD in $\alpha $-quartz.
Platz\H oder \cite{19} has measured the vacuum uv reflectance of $\alpha
$-quartz and has found that the sharp peak at $0.1190\mu m$ is due to
this exciton transition and that the other peaks at $0.105\mu m$,
$0.0840\mu m$ and $0.0710\mu m$ are due to the interband transitions.
The characteristic wave lengths in formulae I~- III and V are located at
the bottom of the dip between two peaks at $0.084$ and $0.105\mu m$. The
bottom is at $0.0940\mu m$ and it is in agreement with ORD formulae. On
the other hand characteristic wave lengths in two~-~termed formulae IV
and VI belong to two different interband transitions. The peak at
$0.105\mu m$ is, in fact, split into three very near peaks. Two bottoms
are located between $0.10$ and $0.11\mu m$. The characteristic wave
lengths in first terms in formulae IV and VI are then very close to
above mentioned values. Both other terms contain $\lambda _{0}$ which is
located at $\approx 0.084\mu m$.

The connection between characteristic wave lengths and interband
transitions is in agreement with the model theory of coupled
oscillators. The normal modes of vibrations are nothing but the
tight-binding exciton model in quantum mechanical term.

Natori \cite{20} has pointed out that coupled oscillator model pays
no attention to the effect of the electron transfer from atom to atom
that gives rise to the ordinary electronic band. For this reason Natori
has discussed ORD in term of the ordinary energy band theory. He has
obtained for $\alpha $-quartz following formula
\begin{eqnarray}
\rho (\omega )&=&911.22 \Bigl(\sqrt{10.834-\hbar \omega
}+\sqrt{10.834+\hbar \omega
}-2\sqrt{10.834}\Bigl)+ \nonumber \\
&&+5708.8 \Bigl(\frac{1}{\sqrt{10.834-\hbar \omega
}}+\frac{1}{\sqrt{10.834+\hbar \omega }}-\frac{2}{\sqrt{10.834}}\Bigl),
\label{25}
\end{eqnarray}
where $\hbar \omega $ is measured in $eV$. In this formula he has taken
into account $11$ points of measurements from $0.6708$ to $0.1525\mu m$
and calculated $\rho $ with the aid of eq. (\ref{25}). But in this region
$g=2.16$ while Chandrasekhar's formula gives $g=0.992$. Moreover it
follows that the deviations $\Delta \rho =\rho _{exp}-\rho _{cal}$ are
of systematic nature when using formula (\ref{25}). In the region from
$0.6708$ to $0.3403\mu m$ $\Delta \rho $ in negative and the deviations
have decreasing tendency from $-0.11$ to zero at $\lambda =0.3403\mu m$.
In the region from $0.3403$ to $0.1525\mu m$ $\Delta \rho $ is positive
with increasing tendency from zero to $+3.9$ at $\lambda =0.1725\mu m$.
The deviation $\Delta \rho $ at $\lambda=0.1525\mu m$ is again negative
but very great, that is $-5.8$. In formula (25) $\hbar \omega
_{0}=10.834eV$. It means that $\lambda _{0}=0.11443\mu m$ which located
it at the bottom of the dip between two peaks at $0.105$ and $0.1190\mu
m$. Natori has pointed out that this location of $\lambda _{0}$ is in
agreement with the used model and that the formula (\ref{25}) is more preferable
than Chandrasekhar's formula. But as we have shown his formula does not
enable good approximation of experimental data.

The analysis cannot stop here but we can conclude that ORD of $\alpha
$-quartz is due to its crystal structure because in all appropriate
formulae Chandrasekhar's terms play a predominant role and just these
terms reflects the optical activity of crystalline origin. It should be
noted that Drude's term in our formula does not reflect optical activity
of molecular origin as judged by some authors (see e. g. \cite{21} and
references in this paper). This term reflects the splitting of rotatory
strengths of normal modes of vibrations (see Appendix). Meanwhile, we do
not know the cause of this splitting but we expect that this splitting
is caused by crystalline field. We hope in the future to apply the model
of coupled oscillators to some real crystals including the effects of
crystalline field.

Some authors \cite{21} ascribe our combined formula to Agranovich
\cite{9,10}.
But Agranovich has pointed out that Drude's term vanishes in his formula
when taking into account crystal consisting from optically inactive
molecules. The difference between our formula and formula of Born
\cite{22}
has been discussed by us in \cite{8}. In the Appendix we shall show that
formula of Agranovich and Tsvirko is special case of our combined
formula as well as formula derived recently by Kato et alii \cite{23}.

\section*{Appendix: Rotational Strengths of Normal Modes of
Vibrations}
\hspace \parindent
A transition is optically active when it has non-vanishing rotational
strengths. For a transition from some ground state $\vert 0\rangle $ to
the excited state $\vert n\rangle $ the rotational strength is a imaginary
part of scalar product of the electric dipole moment $\vec \mu _{e}$ and
the magnetic dipole moment $\vec \mu _{m}$. These moment are $\langle
0\vert \vec \mu _{e}\vert n\rangle$ and $\langle n\vert \vec \mu
_{m}\vert 0\rangle $ so that
\begin{equation}
R_{0n}=\mbox{Im}\langle 0\vert \vec \mu _{e}\vert n\rangle \langle n\vert \vec
\mu _{m}\vert 0\rangle .
\label{26}
\end{equation}

In the case of coupled oscillators the excited state $\vert n\rangle $
is split into two states $\vert 1n\rangle$ and $\vert 2n\rangle$.
Therefore the rotational strength needs be considered between ground state
and both excited states. The total electric moment of compound
oscillator is
\begin{eqnarray}
\vec \mu _{e}&=&e\left(\vec r_{1}+\vec r_{2}\right)= \nonumber \\
&=&e\bigl[\vec{\imath}\alpha r_{1}+\vec{\jmath}\beta r_{1}+
\vec{\imath}\left(\alpha \cos
\theta -\beta \sin \theta \right)r_{2}
+\vec{\jmath}\left(\alpha \sin \theta
+\beta \cos \theta \right)r_{2}\bigl].
\label{27}
\end{eqnarray}

Here $\vec{\imath}$ and $\vec{\jmath}$ are the unit vectors along $x$ and $y$ axis
and $\vec r_{1}$ and $\vec r_{2}$ are position vectors of vibrating
electrons in both single oscillators. Because we study the rotational
strengths for light propagating along $c$-axis we taken into account the
vibrations along $x$ and $y$ axes. Using the normal coordinates
\begin{equation}
r_{1}=\frac{1}{\sqrt{2}}\left(q_{1}+q_{2}\right),\qquad \qquad
r_{2}=\frac{1}{\sqrt{2}}\left(q_{1}-q_{2}\right)
\label{28}
\end{equation}
we obtain from eq. (\ref{27}) the normal components of $\vec \mu _{e}$ as
follows
\begin{equation}
\vec \mu _{e_{q_{1}}}=\frac{e}{\sqrt{2}}\Bigl\{\vec{\imath}\left[\alpha
\left(1+\cos \theta \right)-\beta \sin \theta \right]+\vec{\jmath}\left[\beta
\left(1+\cos \theta \right)+\alpha \sin \theta \right]\Bigl\}q_{1}
\label{29}
\end{equation}
and
\begin{equation}
\vec \mu _{e_{q_{2}}}=\frac{e}{\sqrt{2}}\Bigl\{\vec{\imath}\left[\alpha
\left(1-\cos \theta \right)+\beta \sin \theta \right]+\vec{\jmath}\left[\beta
\left(1-\cos \theta \right)-\alpha \sin \theta \right]\Bigl\}q_{2}.
\label{30}
\end{equation}

The magnetic dipole moment of compound oscillator is
\begin{equation}
\vec \mu _{m}=\frac{e}{2c}\left[\left(\frac{\vec d}{2}\times \dot{\vec
r_{2}}\right)-\left(\frac{\vec d}{2}\times \dot{\vec
r_{1}}\right)\right],
\label{31}
\end{equation}
where $\vec d$ is linear distance between two centers of gravity of both
single oscillators. We can write instead of eq. (\ref{31})
\begin{equation}
\vec \mu _{m}=-\frac{ed}{4c}\left[\vec k\times\left(\dot{\vec
r_{1}}-\dot{\vec r_{2}}\right)\right].
\label{32}
\end{equation}

Here $\vec k$ is the unit vector along $z$-axis which is identical with
crystal $c$-axis. We see that magnetic dipole moment arises from a
particular spatial distribution of coupled oscillators. When $\vec
r_{1}$ and $\vec r_{2}$ colinear the magnetic dipole moment is zeroth.
The normal components of $\vec \mu _{m}$ are
\begin{equation}
\vec \mu _{m_{q_{1}}}=-\frac{ed}{\sqrt{32}c}\Bigl\{\vec{\jmath}\left[\alpha
\left(1-\cos \theta \right)+\beta \sin \theta \right]-\vec{\imath}\left[\beta
\left(1-\cos \theta \right)-\alpha \sin \theta \right]\Bigl\}\dot q_{1}
\label{33}
\end{equation}
and
\begin{equation}
\vec \mu _{m_{q_{2}}}=-\frac{ed}{\sqrt{32}c}\Bigl\{\vec{\jmath}\left[\alpha
\left(1+\cos \theta \right)-\beta \sin \theta \right]-\vec{\imath}\left[\beta
\left(1+\cos \theta \right)+\alpha \sin \theta \right]\Bigl\}\dot q_{2}
.
\label{34}
\end{equation}

According to eq. (\ref{26}) we get
\begin{equation}
R_{1_{n0}}=\mbox{Im}\left[-\frac{2e^{2}d\left(\alpha ^{2}+\beta ^{2}\right)\sin
\theta \langle 0\vert q_{1}\vert 1n\rangle \langle 1n\vert \dot
q_{1}\vert 0\rangle }{8c}\right]
\label{35}
\end{equation}
and
\begin{equation}
R_{2_{n0}}=\mbox{Im}\left[\frac{2e^{2}d\left(\alpha ^{2}+\beta ^{2}\right)\sin
\theta \langle 0\vert q_{2}\vert 2n\rangle \langle 2n\vert \dot
q_{2}\vert 0\rangle }{8c}\right].
\label{36}
\end{equation}

The matrix elements $\langle \eta n\vert \dot q_{\eta }\vert 0\rangle$
may be written as follows
\begin{equation}
\langle \eta n\vert \dot q_{\eta }\vert 0\rangle =i\omega _{\eta
n0}\langle \eta n\vert q_{\eta }\vert 0\rangle
\label{37}
\end{equation}
and using definition of oscillator strength
\begin{equation}
f_{q_{\eta }}=\frac{2m\omega _{\eta n0}\vert \langle \eta n\vert q_{\eta
}\vert 0\rangle \vert ^{2}}{\hbar}
\label{38}
\end{equation}
we obtain for $R_{1_{n0}}$ and $R_{2_{n0}}$
\begin{equation}
R_{1_{n0}}=-\frac{\hbar e^{2}d\left(\alpha ^{2}+\beta
^{2}\right)f_{q{1}}\sin \theta }{8mc}
\label{39}
\end{equation}
and
\begin{equation}
R_{2_{n0}}=\frac{\hbar e^{2}d\left(\alpha ^{2}+\beta
^{2}\right)f_{q{2}}\sin \theta }{8mc}.
\label{40}
\end{equation}

We see that $R_{1_{n0}}$ and $R_{2_{n0}}$ substituted into eq. (\ref{12}) gives
directly eq. (\ref{9}) and from eq. (\ref{13}) we get eq. (\ref{10}).

In linear harmonic oscillator approximation one may evaluate the matrix
elements in eq. (\ref{38}), that is
\begin{equation}
\vert \langle \eta n\vert q_{\eta }\vert 0\rangle \vert
^{2}=\frac{\hbar }{2m\omega _{\eta _{n0}}}
\label{41}
\end{equation}
and from thus $f_{q_{1}}=f_{q_{2}}=1$. Then Drude's term vanishes in eq.
(10) and we get Chandrasekhar's formula. Of course, in real molecules
$f_{q_{1}}=f_{q_{2}}\leq n$, where $n$ is a number of oscillating
electrons. In Heitler~-~London approximation one may write $\langle
1n\vert q_{1}\vert 0\rangle =\langle 2n\vert q_{2}\vert 0\rangle $ and
therefore
\begin{equation}
\frac{f_{q_{1}}}{\omega _{1_{n0}}}=\frac{f_{q_{2}}}{\omega
_{2_{n0}}}=\frac{f_{0}}{\omega _{n0}}.
\label{42}
\end{equation}

Here $f_{0}$ is, as before, average value of both oscillator strengths,
that is oscillator strength of single oscillator. Further $\omega _{n0}$
is characteristic frequency of uncoupled oscillator. Using eq. (\ref{7}) we
get from eq. (\ref{42})
\begin{equation}
f_{q_{1}}+f_{q_{2}}=2f_{0},\qquad \qquad f_{q_{2}}-f_{q_{1}}=-\frac{2\pi
^{2}f_{0}\epsilon }{\omega _{n0}^{2}}=-\frac{\epsilon f_{0}\lambda
_{0}^{2}}{2c^{2}}
\label{43}
\end{equation}
and substituting these expressions into eq. (\ref{10}) one has
\begin{equation}
\rho (\lambda )=\frac{\pi Ne^{2}d\left(\alpha ^{2}+\beta
^{2}\right)\epsilon f_{0}\lambda _{0}^{4}\sin \theta
}{mc^{4}}\cdot\frac{\lambda ^{2}+\lambda _{0}^{2}}{\left(\lambda
^{2}-\lambda _{0}^{2}\right)^{2}}
\label{44}
\end{equation}
or
\begin{equation}
\rho (\lambda )=\frac{K^{(3)}\left(\lambda ^{2}+\lambda
_{0}^{2}\right)}{\left(\lambda ^{2}-\lambda _{0}^{2}\right)^{2}}.
\label{45}
\end{equation}

This formula is identical with that obtained by Agranovich \cite{9,10} and
by Tsvirko \cite{11}. We see that formula (\ref{45}) is nothing but the
special case of our combined formula (\ref{3}).

Kato et alii \cite{23} have derived ORD formula from the exciton dispersion
theory. Their formula is
\begin{equation}
\rho (\lambda )=\frac{K^{(4)}\left(3\lambda ^{2}-\lambda
_{0}^{2}\right)}{\left(\lambda ^{2}-\lambda _{0}^{2}\right)^{2}}.
\label{46}
\end{equation}

Without entering into the details of the microscopic theory it is
interesting that when $K^{(2)}=2K^{(1)}$ in our formula (\ref{3}) this formula
reduces to formula (\ref{46}). On the other hand formula (\ref{4}) may be
formally obtained from formula (\ref{3}) when $2K^{(1)}=-K^{(2)}$
but it is nothing but
Heitler~-~London approximation discussed in eqs. (\ref{42})~-~(\ref{45}).

\label{vys1-kon}


\vspace{1cm}


\begin{thebibliography}{99}
\bibitem{1} Burkov, V. I., Ivanov, V. V., Kizel, V. A., Semin, G. S.: Works
of Moscow Physico-technical Institute {\bf 5}, 1974, 4.
\bibitem{2} Burkov, V. I., Ivanov, V. V., Kizel, V. A., Semin, G. S.: Optika i
Spektroskopiya {\bf 37}, 1974, 740.
\bibitem{3} Duclaux, J., Jeantet, P.: J. Phys. {\bf 7}, 1926, 200.
\bibitem{4} Lowry, T. M., Coode-Adam, W. R.: Phil. Trans. {\bf A266}, 1927, 291.
\bibitem{5} Servant, R.: Ann. Phys. {\bf 12}, 1939, 397.
\bibitem{6} Chandrasekhar, S.: Proc. Ind. Acad. Sci. {\bf A~35},
1952, 103.
\bibitem{7} Chandrasekhar, S.: Proc. Roy. Soc. {\bf A259}, 1961, 531.
\bibitem{8} Vy\v s\'\i n, V.: Proc. Phys. Soc. {\ 87}, 1966, 55.
\bibitem{8a} Nelson, D. F.: JOSA {\bf B6}, 1989, 1110.
\bibitem{9} Agranovich, V. M.: Optika i Spektroskopiya {\bf 1}, 1956,
338.
\bibitem{10} Agranovich, V. M.: Optika i Spektroskopiya {\bf 2}, 1957,
738.
\bibitem{11} Tsvirko, J. A.: Zhurnal Eksp. i Teor. Fiz. {\bf 38}, 1960,
1615.
\bibitem{12} Katzin, L. I.: J. Phys. Chem. {\bf 68}, 1964, 2367.
\bibitem{13} B\H urer, T., Katzin, L. I.: J. Inorg. Nucl. Chem. {\bf
29}, 1967, 2715.
\bibitem{14} Hennessey, P., Vedam, K.: JOSA {\bf 65}, 1975, 436.
\bibitem{15} Vy\v s\'\i n, V.: Optics Comm. {\bf 1}, 1970, 307.
\bibitem{16} Jank\accent23u, V.: Optica Acta {\bf 16}, 1969, 225.
\bibitem{17} Brill, R., Hermann, C., Peters, V.: Ann Phys. {\bf 41},
1942, 233.
\bibitem{18} Young, R. A., Post, B.: Acta Crystallogr. {\bf 15}, 1962,
337.
\bibitem{19} Platz\H oder, K.: Phys. Stat. Sol. {\bf 29}, 1968, K63.
\bibitem{20} Natori, K.: Thesis of Faculty of Sciences. Tokyo, University of
Tokyo 1973.
\bibitem{21} Kizel, V. A., Krasilov, J. I., Burkov, V. I.: Uspekhi Fiz. Nauk
{\bf 114}, 1974, 295.
\bibitem{22} Born, M., G\H oppert-Mayer, M.: Handbuch d. Phys.
24/2. Berlin, Springer-Verlag 1933.
\bibitem{23} Kato, T., Tsujikawa, I., Murao, T.: J. Phys. Soc. Japan {\bf 14},
1973, 763.


\end{thebibliography}
\end{document}